# Minimizing the scattering of a non-magnetic cloak

Jingjing Zhang[1*], Yu Luo[2], and Niels Asger Mortensen[1]

[1] *DTU Fotonik - Department of Photonics Engineering, Technical University of Denmark, DK-2800 Kongens Lyngby, Denmark*

[2] *The Blackett Laboratory, Department of Physics, Imperial College London, London SW7 2AZ, UK*

**Abstract**

Nonmagnetic cloak offers a feasible way to achieve invisibility at optical frequencies using materials with only electric responses. In this letter, we suggest an approximation of the ideal nonmagnetic cloak and quantitatively study its electromagnetic characteristics using a full-wave scattering theory. It is demonstrated that the forward scattering of the impedance matched cloak increases dramatically as the thickness of the cloak decreases. Nevertheless, it is still possible to effectively reduce the total scattering cross section with a very thin cloak whose impedance is not matched to the surrounding material at the outer boundary. Our analysis also provides the flexibility of reducing the scattering in an arbitrary direction.

[*] *Author to whom correspondence should be addressed; electronic mail: jinz@fotonik.dtu.dk*

Realizing invisibility has been a dream of people for centuries. Recently, this topic has intrigued increasing interests [1-20], especially since the idea of designing invisibility cloaks with transformation methodology was proposed [1-3]. Inspired by those theoretical studies, a cylindrical cloak at microwave frequency was soon experimentally realized [4], which was based on a set of simplified cloak parameters [5]. However, as pointed out in Ref. [6], the design for the microwave cloak is not directly adaptable to optical frequencies where the concept of 'invisibility' would have far reaching potential. Ref. [6] gave a prescription for a nonmagnetic cloak working at the optical frequency range and an experiment on the optical cloak was later reported [7]. The material parameters for those approximate nonmagnetic optical cloaks, originated from a linear transformation function [5], have a significant limitation that the impedance of the cloak is strongly mismatched at the outer surface. As an improvement approach, a high-order coordinate transformation, which renders the cloak perfectly matched to the free space, is made use of to eliminate the undesired scattering from the cylindrical nonmagnetic cloak [8, 9]. And Ref. [10] considers the boundary conditions at the inner boundary of the high-order point transformed cloak. However, it has been demonstrated theoretically that the drawback of the simplified cloak is more than just non-zero reflectance at the boundary [11]. Several alternative approaches have been considered to overcome this obstacle [12, 13].

In this letter, we suggest a generalized form of simplified nonmagnetic cloak parameters, which can be optimized at a given frequency to minimize the scattering as desired. With a full wave analysis process, we theoretically demonstrate that the cloak system with a matching outer surface may have a large total scattering cross section when the cloak layer is kept thin, while it truly performs well for the backscattering detection. This method gives the freedom of designing nonmagnetic cloaks with small thicknesses, which are nearly transparent for detections from arbitrary given directions.

For a transverse-magnetic (TM) incidence, we can make an approximation for the two-dimensional ideal cloak while sustaining the important nonmagnetic feature as, $\varepsilon_{\rho\rho} = (f(\rho)/\rho)^2$, $\varepsilon_{\varphi\varphi} = (f'(\rho))^2$, $\mu_{zz} = 1$. Here $f(\rho)$ is an arbitrary function which satisfies $0 \leq f(\rho) \leq R_2$ and $R_1 \leq \rho \leq R_2$. The corresponding wave equation becomes

$$\frac{1}{\rho}\frac{\partial}{\partial \rho}\frac{\rho}{[f'(\rho)]^2}\frac{\partial}{\partial \rho}H_z + \frac{1}{f^2(\rho)}\frac{\partial^2}{\partial \varphi^2}H_z + k_0^2 H_z = 0. \tag{1}$$

In order to make the material parameters amenable to analytic solution of Eq. (1) and the boundary conditions $f(R_2) = R_2$ and $f(R_1) = 0$, we restrict ourselves to a transformation function expressed as

$$f(\rho) = R_2 \left(\frac{\rho^2 - R_1^2}{R_2^2 - R_1^2}\right)^{\frac{1}{1+\alpha}}. \tag{2}$$

where $\alpha$ is an adjustable parameter. The general solution of Eq. (1) takes the form $H_z = [k_0 f(\rho)]^{\frac{1-\alpha}{2}} B_\nu [k_0 f(\rho)] e^{in\varphi}$, where $k_0 = \omega\sqrt{\mu_0 \varepsilon_0}$, while $B_\nu$ represents a Bessel function of $\nu$-th order, where $\nu = \sqrt{n^2 + (\alpha-1)^2/4}$.

Suppose a TM polarized plane wave with the unit amplitude $H_z^i = e^{ik_0 x}$ is incident upon the cloak along the *x* direction. The incident fields, scattered fields, and the fields inside the cloak layer are then conveniently written in polar coordination as

$$H_z^i = \sum_{n=-\infty}^{\infty} a_n J_n(k_0\rho)\cos n\varphi, \tag{3a}$$

$$H_z^s = \sum_{n=-\infty}^{\infty} b_n H_n^{(1)}(k_0\rho)\cos n\varphi, \tag{3b}$$

$$H_z^c = [k_0 f(\rho)]^{\frac{1-\alpha}{2}} \sum_{n=-\infty}^{\infty} \{c_n J_\nu[k_0 f(\rho)] + d_n N_\nu[k_0 f(\rho)]\}\cos n\varphi. \tag{3c}$$

Here $a_n=i^n$, $b_n$, $c_n$, $d_n$ are all unknown expansion coefficients to be determined. By applying the continuities of $H_z$ and $E_\varphi$ at $R_2$ and $R_1$ [The inner boundary of the cloak is assumed to be a perfect electric conductor (PEC)] and considering Eq. (2), all the undetermined coefficients can then be obtained as

$$b_n = \frac{B J_n(k_0 R_2) - A J_n'(k_0 R_2)}{A H_n^{(1)\prime}(k_0 R_2) - B H_n^{(1)}(k_0 R_2)} a_n, \tag{4a}$$

$$c_n = \frac{J_n(k_0 R_2) H_n^{(1)\prime}(k_0 R_2) - J_n'(k_0 R_2) H_n^{(1)}(k_0 R_2)}{A H_n^{(1)\prime}(k_0 R_2) - B H_n^{(1)}(k_0 R_2)} a_n, \tag{4b}$$

$$d_n = 0. \tag{4c}$$

where $A = (k_0 R_2)^{\frac{1-\alpha}{2}} J_\nu(k_0 R_2)$ and $B = (k_0 R_2)^{-\frac{1+\alpha}{2}}\left[(1-\alpha)J_\nu(k_0 R_2)/2 + k_0 R_2 J_\nu'(k_0 R_2)\right]$. $J_n$, $N_n$, and $H_n^{(1)}$ represent the nth order of Bessel function, Neumann function, and Hankel function of the first kind, respectively. We notice that $b_n$ is in general non-zero, indicating the existence of scattering outside the cloak.

In what follows, we aim to discuss the way of finding the value of the variable $\alpha$ which can optimize the performance of the cloak. As is commonly known, the impedance matching is a necessary condition for the reflection reduction. Hence we first take into account the condition that the impedance of the cloak is matched to the free space at the outer boundary ($\varepsilon_\varphi|_{\rho=R_2} = [f'(R_2)]^2 = 1$). With Eq. (3), we straightforwardly have $\alpha = (R_2^2 + R_1^2)/(R_2^2 - R_1^2)$.

In order to study the performance of this cloak in terms of the thickness of the coating layer, we fix the inner boundary of the cloak to be $R_1$=1.5μm, and plot the normalized total cross section [21] as a function of the outer boundary, as shown in Fig. 1. The source works at 200THz in all the calculations. The total scattering cross section increases dramatically as the thickness of the cloak decreases. The simplified cloak show the effectiveness of reducing the scattering only when $R_2 \Box R_1$. The insets display the electric field distributions for two cloaks where $R_2 = 3R_1$ and $R_2 = 4/3R_1$. It can be clearly observed that when $R_2$ is three times larger than $R_1$, the simplified cloak shows a good performance comparable to that of an ideal cloak. However, a shadow is formed in the forward path of the cloak when the outer radius is close to the inner radius.

Although the impedance matching condition provides a solution to the determination of parameters if the thickness of the cloak is not a significant factor, it is still necessary to find a general optimization method which is applicable to a cloak with thin coating layer. Here we study this problem by fixing both the inner and outer boundaries of the cloak and calculating the scattering cross sections in different directions in terms of $\alpha$. We consider cloaks with three different thicknesses (case 1: $R_2$=5μm; case 2: $R_2$=4μm; case 3: $R_2$=3μm; $R_1$=2.5μm for all cases) and calculate the normalized total scattering cross sections [see Fig. 2 (a)]. By comparing the three curves, we find that as the thickness of the cloak reduces, the $\alpha$ which leads to a minimum total scattering cross section deviates more and more from the $\alpha$ which makes the simplified cloak matched to the free space. The curves also imply the way of minimizing the total scattering cross section. As a reference, we also plot the normal total scattering cross sections for the PEC scatterers with radii $R= 2.5\mu$m and $R=5\mu$m. It is worth noticing that when we choose certain values of $\alpha$, the corresponding cloaks themselves will induce such large scattering that the total scattering cross sections are even larger than that of the PEC core. For comparison, we also calculate the normalized forward scattering cross sections [20] in terms of $\alpha$ to study the bistatic (transmitter and receiver in different locations) scattering properties of cloaks with different thicknesses.

Sometimes, for a monostatic detection where the transmitter and the receiver lie in the same position, we only need to consider the backscattering property of the cloak. In order to determine the optimized parameters in this condition, we plot the normalized backward scattering cross section in terms of $\alpha$, as shown in Fig. 2 (c). For different thickness cases, we find that the cloaks with matching outer boundaries are almost invisible for monostatic detections.

In fact, for detector located in any other direction, we can plot the scattering cross section curve similarly and identify a certain $\alpha$ which can minimize the scattering. For the purpose of a clearer illustration, we plot the scattering pattern of a cloak whose outer and inner boundaries are 3μm and 2.5μm, respectively. The black solid line denotes the case where the $\alpha$ is picked at the matched point, and the red solid line corresponds to the $\alpha$ which makes a minimum total scattering cross section.

Although the impedance matched cloak shows satisfying performance in a wide range of degrees in the backward side, it will induce a strong scattering in the forward directions, which is even larger than that caused by the PEC obstacle (denoted by cyan solid line). The near field distributions for the corresponding matched cloak ($\alpha = 5.5$) and cloak with minimum total scattering cross section ($\alpha = 1.8$) are displayed in panels (b) and (c), respectively. By comparing Fig. 3 (b) and (c), we can see that the forward scattering reduction property can be distinctly improved by choosing the $\alpha$ corresponding to the "minimum point" other than the "matched point".

In conclusion, we have analytically studied the performance of nonmagnetic cloaks with reduced material parameters. It is demonstrated that although the boundary impedance matching condition can render a cloak with good backscattering performance, it does not provide a recipe for reducing scattering in other orientations. As a solution, a parameter optimization scheme is suggested, which makes it possible to design cloaks of required scattering capability. Our method provides the flexibility of designing nonmagnetic cloaks for both monostatic and bistatic detections.

The work is supported by the Hans Christian Ørsted postdoctoral fellowship. Jingjing Zhang would like to appreciate the discussions with Prof. Hongsheng Chen.

**Figure captions**

Fig. 1 (color online) Normalized total scattering cross sections in terms of $R_2/R_1$, where $R_1$ is fixed to be 1.5μm. The insets show the field distributions for the cases where $R_2 = 4/3R_1$ and $R_2 = 3R_1$.

Fig. 2 (color online) (a) Normalized total scattering cross sections in terms of $α$ for cloaks of different thicknesses and PEC scatters of different radii. (b) Normalized forward scattering cross sections for cloaks of different thicknesses and the PEC scatterer. (c) Normalized backward scattering cross sections for cloaks of different thicknesses and the PEC scatterer.

Fig. 3 (color online) (a) Far filed scattering patterns for cloaks and the PEC scatterer. The cyan solid line corresponds to a PEC scatter with the radius of 2.5μm. The black solid line corresponds to the cloak ($R_1$=2.5μm, $R_2$=3μm) with a matched outer boundary. The red solid line corresponds to the cloak (the same thickness) with the minimum total scattering cross section. Panels (b) and (c) show the near field properties for the aforementioned two cases, respectively.

Fig. 1

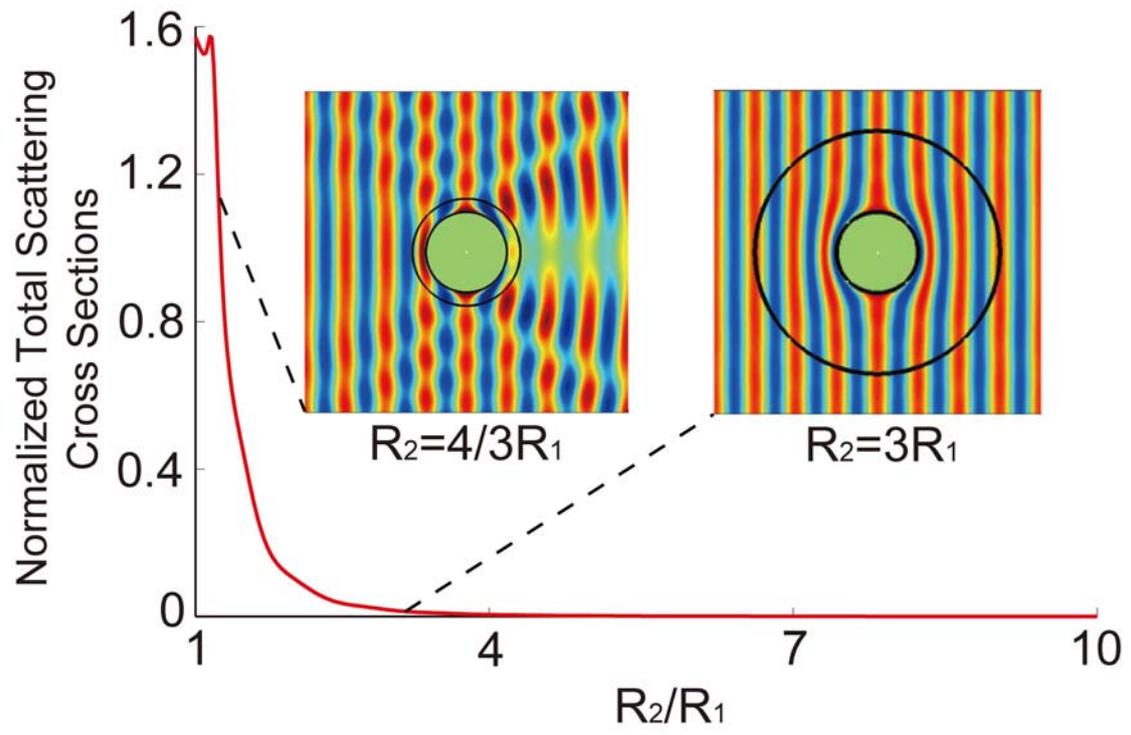

Fig. 2

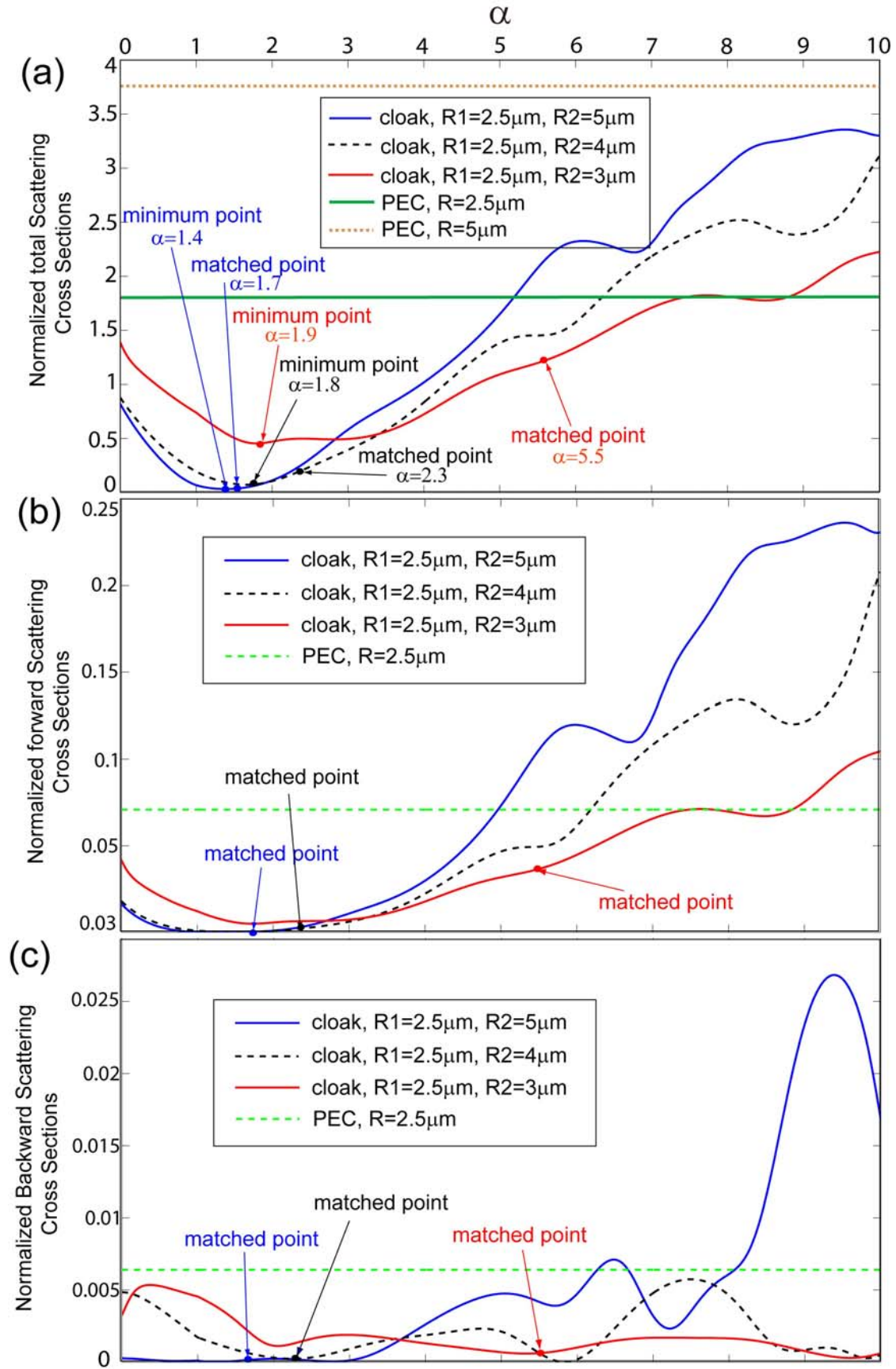

Fig. 3

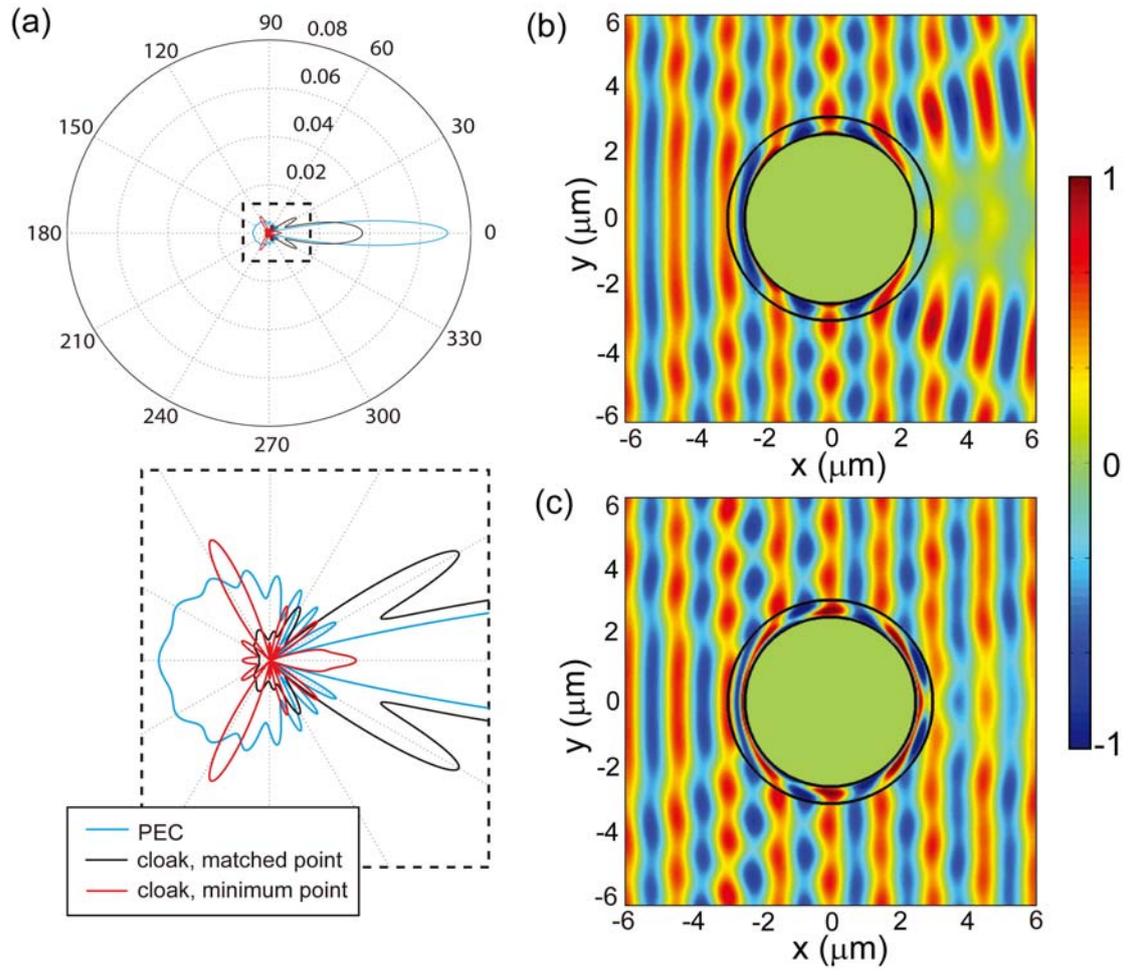